\documentclass[12pt]{article}
\usepackage{graphicx}
\usepackage{amsmath}
\usepackage{hyperref}
\usepackage{multirow}
\usepackage{booktabs}
\usepackage{INTERSPEECH2021}
\usepackage{graphicx}

\title{Speaker Emotion Recognition: Leveraging Self-Supervised Models for Feature Extraction Using Wav2Vec2 and HuBERT}

\name{Pourya Jafarzadeh$^1$, Amir Mohammad Rostami$^1$, Padideh Choobdar}

\address{
  $^1$Lab of Artin, TOSAN TECHNO}
\email{[p.jafarzadeh, am.rostami]@tosantechno.com, p.choobdar@gmail.com}

\date{}

\begin{document}

\maketitle

\begin{abstract}
Speech is the most natural way of expressing ourselves as humans. Identifying emotion from speech is a nontrivial task due to the ambiguous definition of emotion itself. Speaker Emotion Recognition (SER) is essential for understanding human emotional behavior. The SER task is challenging due to the variety of speakers, background noise, complexity of emotions, and speaking styles. It has many applications in education, healthcare, customer service, and Human-Computer Interaction (HCI). Previously, conventional machine learning methods such as SVM, HMM, and KNN have been used for the SER task. In recent years, deep learning methods have become popular, with convolutional neural networks and recurrent neural networks being used for SER tasks. The input of these methods is mostly spectrograms and hand-crafted features. In this work, we study the use of self-supervised transformer-based models, Wav2Vec2 and HuBERT, to determine the emotion of speakers from their voice. The models automatically extract features from raw audio signals, which are then used for the classification task. The proposed solution is evaluated on reputable datasets, including RAVDESS, SHEMO, SAVEE, AESDD, and Emo-DB. The results show the effectiveness of the proposed method on different datasets. Moreover, the model has been used for real-world applications like call center conversations, and the results demonstrate that the model accurately predicts emotions.
\end{abstract}

\begin{figure*}[ht]
    \centering
    \includegraphics[width=0.7\textwidth]{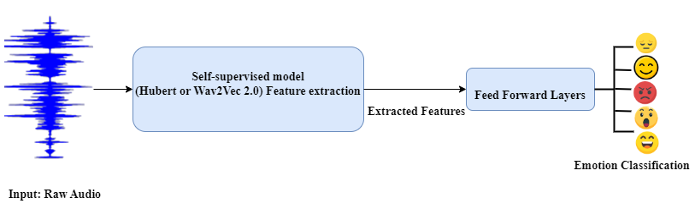}
    \caption{The high-level description of our model. The input is a raw audio signal, in the second stage, features are extracted by transformer based model. Then, two feed forward layers classify the signal.}
    \label{fig:hld}
\end{figure*}

\section{Introduction}
Speech is one of the primary ways of expressing emotions. Therefore, a system that can recognize, interpret, and respond to the emotions expressed in speech is highly valuable. Emotions influence both the vocal characteristics and the linguistic content of speech. In recent years, enormous efforts have been devoted to developing methods for automatically identifying human emotions from speech signals, a field known as speaker emotion recognition (SER). SER systems have many applications including human-machine interactions \cite{1,2}, medicine \cite{3}, and psychology \cite{4}. Moreover, evaluation of the conversations in a call centre is one of the major applications of SER, the manager of a call centre can analyse the performance of the operators. The operators are able to consider the feeling of the customers during conversation to handle necessary cases.

In all of the classification problems one of the main step is feature extraction. Before using deep learning most of the conventional machine learning methods used hand-crafted features speech recognition including Mel-frequency cepstral coefficients (MFCCs, pitch, zero-crossing, Fourier transform, and energy of signals \cite{5,6,7,8,9,10}. After growing deep learning usage in speech processing tasks, the models extracted features automatically by deep-based models. They used convolutional neural networks (CNNs) to extract spatial features from input, the input was mostly spectrograms. Also, they used recurrent neural networks (RNNs) to evaluate temporal features. The models extracted local information and long-term contextual dependencies \cite{11,12,13,14,15,16,17,18,19,20}. Additionally, the attention mechanism is another popular method that has been used in different fields such as speech recognition \cite{18} , visual object classification \cite{19} and document classification \cite{20}. It is a powerful concept that helps models focus on specific parts of the input sequence when generating each part of the output sequence. In the SER problem, models try to ignore silence frames and other parts of the utterance that do not carry emotional content \cite{21,22,23,24}.
Our contribution in this study lies in solving speaker emotion recognition by using transformer-based models, HuBERT \cite{25} and Wav2vVec \cite{26}, which trained in a self-supervised method on the huge amount of data. These models are very powerful for feature extraction, so we used the models to extract features from raw audio data. We will delve deeper into the models in the next sections.

\section{Related Works}

Different machine learning algorithms have been used to construct a good classifier for emotion classification. As mentioned, conventional machine learning methods such as support vector machines (SVM) \cite{27,28,10,29}, k-nearest neighbours (KNN) \cite{30}, and decision tree \cite{31} have been used for emotion recognition. In \cite{30}, two utterances were aligned in the emotion space defined by emotograms using Multi-Dimensional Dynamic Time Warping (MD-DTW) \cite{32}. A KNN classifier was then applied to assign a final emotion class label based on the MD-DTW measure. KNN assigns a label to a test utterance based on the labels of its k nearest neighbors, with the final label determined by a majority vote among the neighbors.
 Each of the classifiers has its own advantages and disadvantages. These methods mainly use hand-crafted features such as pitch, energy, zero-crossing rate, Mel-filterbank features, and MFCCs. These are often referred to as Low-Level Descriptors (LLD). Since MFCC models the human auditory perception system, it is the most popular spectral feature. Moreover, CNN-based models take the signal's spectrogram as input, which is a visual representation of the spectrum of frequencies of a signal as it varies with time \cite{14,15}. In \cite{14}, spectrograms were generated from the input speech signal, followed by the use of three convolutional layers and three fully connected layers to extract features from the spectrogram images. A spectrogram is a visual representation of signal strength over time at various frequencies present in a waveform. It is depicted as a two-dimensional graph, with time on the horizontal axis, frequency on the vertical axis, and the amplitude of the frequency components at a specific time represented by the intensity or color at each point. The spectrogram is computed from the speech signal by applying the Fast Fourier Transform (FFT), resulting in a time-frequency representation.
Some methods apply CNNs to extract local features from the input, then use LSTM layers to learn long-term dependencies from the extracted local features \cite{13,21}.  In \cite{13}, two CNN-LSTM networks were constructed: one 1D CNN-LSTM network and one 2D CNN-LSTM network, designed to learn local and global emotion-related features from speech and log-mel spectrograms, respectively. Both networks share a similar architecture, consisting of four Local Feature Learning Blocks (LFLBs) and an LSTM layer. Each LFLB, which primarily includes one convolutional layer and one max-pooling layer, is designed to capture local correlations while extracting hierarchical features. The LSTM layer is then used to learn long-term dependencies from these local features. These networks, which combine the strengths of CNNs and LSTMs, capitalize on the advantages of both architectures while mitigating their individual limitations. RNNs, however, are inherently sequential models that do not allow parallelization of their computations, this bottleneck is particularly evident when processing large datasets with long sequences \cite{33}.
The authors in \cite{11,12} have found that spectrograms are more effective than LLD features because LLD features are already decorrelated. One of the advantages of deep-based models is feature extraction; they can extract useful and important features during the training stages. In \cite{11}, the authors analyzed speech using spectrograms and proposed rectangular kernels of varying shapes and sizes, combined with max pooling in rectangular neighborhoods, to extract discriminative features. This approach effectively learns discriminative features from the speech spectrograms.

 The mentioned features are affected by noise and environmental changes, which have a direct impact on the performance of the system. For example, most datasets are recorded in studios by actors, resulting in high-quality audio. If we train a model on this data, it may not perform well in real-world applications due to noise. It would be better to extract more robust features from the audio file. To tackle these problems, we suggest using transformer-based models to extract features from raw audio. These models are trained on a huge amount of data using self-supervised methods, allowing them to learn useful features from raw audio due to the self-attention mechanism in transformers

\section{Proposed Methodology}

A key element of emotion recognition is the ability to extract more distinguishing speech features. In the proposed framework, features are extracted by transformers instead of traditional feature engineering techniques. In this study HuBERT and Wav2Vec2.0 have been used for feature extraction. Both models are transformer-based, and they extract both acoustic features and language modelling from raw audio signals in their architectures. In the following sections we go into more details about two models.
\subsection{HuBERT}
\begin{figure}
    \centering
    \includegraphics[width=0.4\textwidth]{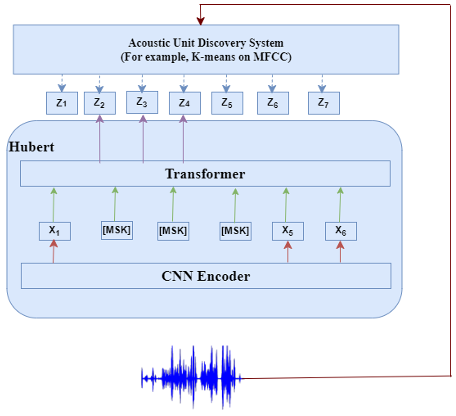}
    \caption{Hubert model architecture}
    \label{fig:hubert}
\end{figure}

Hidden-Unit BERT (HuBERT) provides aligned target labels for BERT-like prediction losses using an offline clustering step (figure \ref{fig:hubert}). It is proposed to overcome three distinct problems in the self-supervised approach: the presence of multiple sound units per input utterance, the absence of lexicon during the pre-training phase, and the absence of explicit segmentation of sound units \cite{25}. In HuBERT, prediction loss is applied to masked regions, forcing a combined acoustic and language model to be learned for unmasked inputs. The model is pre-trained on the Librispeech \cite{34} (960h) and Libri-light \cite{35} (60Kh) benchmarks and it outperformed state-of-the-art methods, it presented in three different models three model sizes pre-trained with HuBERT: BASE (90M parameters), LARGE (300M), and X-LARGE (1B).
The authors in \cite{25} made two decisions regarding mask prediction: how to mask and where to apply the prediction loss. For the first decision,p\% of the timesteps are randomly selected as start indices and spans of l steps are masked using the same strategies as SpanBERT \cite{36}  and wav2vec 2.0 \cite{26}. To address the second decision, cross-entropy loss computed over masked and unmasked time steps asLm andLu. Similarly to language modelling, if the loss is computed only over masked time steps, the model must predict the targets for the unseen frames based on the context. It forces the model to learn both the acoustic representation of unmasked segments and the long-range temporal structure of speech data.

\subsection{Wav2Vec 2.0}
\begin{figure}
    \centering
    \includegraphics[width=0.4\textwidth]{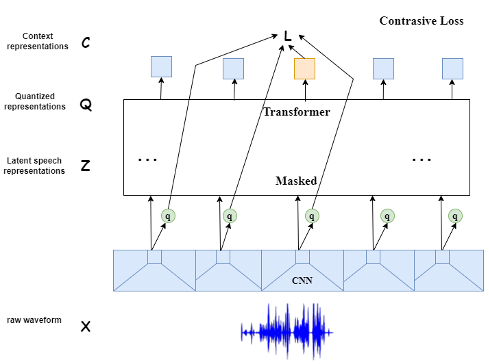}
    \caption{Wav2Vec2.0 model architecture}
    \label{fig:wav2vec}
\end{figure}

The model is a transformer-based model trained with a self-supervised method to extract features from raw audio signals \cite{26} (figure \ref{fig:wav2vec}). It operates by first converting raw audio waveforms into meaningful feature representations using a CNN and a GELU activation function \cite{37} to capture latent speech representations $z_1,z_2,\ldots,z_T$ for $T$ time steps. These initial features are then fed into a transformer network \cite{38}, which is trained using a contrastive loss objective to differentiate between correct and incorrect quantized representations of the audio signal \cite{39}. Using this training method, Wav2Vec 2.0 can learn rich, contextualised embeddings from unlabeled speech data, effectively building context representations over continuous speech and capturing dependencies over the entire sequence of latent representations.
We utilised the pre-trained Wav2Vec 2.0 model to extract high-level features from raw audio inputs, which were then used as inputs for our subsequent processing stages. This approach not only reduces the need for hand-engineered features but also leverages the powerful representations learned by Wav2Vec 2.0, leading to improved performance in our specific application.

\subsection{Model architecture}
The input to our model is a raw audio file, which is sent to the feature extractor module (HUBERT or Wav2Vec 2.0). After extracting the features, we have a matrix with the size of $n\times768$. Depending on the length of the input audio, the length varies. The longer the audio file, the more features we have with the size of 768. In order to project the output of the feature extractor, we calculated the mean of the features across each row, so we will have a vector with a size of 768. Then, we apply two feed-forward layers for classification. The size of the final layer is equal to the number of classes that are compatible with the dataset’s classes (figure \ref{fig:hld}). In the following section, we will report the performance of the proposed model on reputable datasets.
\section{Experimental Results}
We performed SER on five reputable datasets that encompass different languages and emotions to assess the performance of transformer-based models.

\subsection{Datasets}
There are 1440 speech audio samples in the RAVEDESS \cite{40} dataset, recorded by 24 professional actors. Two semantically neutral US English phrases are read while revealing eight emotions (neutral, calm, happiness, sadness, anger, fear, disgust, surprise).
AESDD (Actuated Emotional Speech Dynamic Database) \cite{41,42} is a publicly available dataset for speech emotion recognition. There are utterances of acted emotional speech in the Greek language. There are approximately 500 utterances from five actors. The database contains utterances with five emotions: anger, disgust, fear, happiness, and sadness.
EMODB \cite{43} is a German database with high-quality audio recordings from 10 actors (5 male and 5 female) who produce 10 German utterances (5 short and 5 long sentences) with 7 emotions (anger, neutral, anger, boredom, happiness, sadness, disgust). Some emotional expressions have two versions recorded by the same author. Thus, the database provides about 535 sentences.
The Surrey Audio-Visual Expressed Emotion (SAVEE) \cite{44} database has been recorded as a prerequisite for the development of an automatic emotion recognition system. The database consists of recordings from 4 male actors of 7 different emotions (anger, disgust, fear, happiness, sadness, surprise, and neutral), and 480 British English utterances in total. The sentences were chosen from the standard TIMIT corpus and phonetically balanced for each emotion. The data were recorded in a visual media lab with high-quality audio-visual equipment, processed and labelled.
The SHEMO \cite{45} dataset includes 3000 semi-natural utterances, equivalent to 3 h and 25 min of speech data extracted from online radio plays. The dataset covers speech samples of 87 native-Persian speakers for five basic emotions including anger, fear, happiness, sadness, and surprise, as well as a neutral. It should be noted that we excluded the fear utterances from our test due to their small number in the database, as was done by the authors in \cite{45}.

\subsection{Experiments and Results}
\begin{table*}[ht]
\caption{Comparison of accuracy between proposed models and other methods on Datasets.}
\begin{tabular}{|c|c|c|c|lllllllllllll}
\cline{1-4}
\textbf{Dataset}                   & \textbf{Model}                                                   & \textbf{Weighted Accuracy} & \textbf{Unweighted Accuracy} &  &  &  &  &  &  &  &  &  &  &  &  &  \\ \cline{1-4}
                                   & \textbf{Proposed Method (Hubert)}                                & \textbf{91.66 ± 1.2}       & \textbf{90.48 ± 1.1}         &  &  &  &  &  &  &  &  &  &  &  &  &  \\ \cline{2-4}
                                   & Proposed Method (Wav2Vec2)                                       & 83.34 ± 1.3                & 80.95 ± 1.1                  &  &  &  &  &  &  &  &  &  &  &  &  &  \\ \cline{2-4}
                                   & Ensemble softmax regression \cite{46}                             & 51.50                      & -                            &  &  &  &  &  &  &  &  &  &  &  &  &  \\ \cline{2-4}
                                   & eGeMAPs feature set {\cite{47}}                                     & 42.40                      & -                            &  &  &  &  &  &  &  &  &  &  &  &  &  \\ \cline{2-4}
                                   & DCNN + CFS + MLP {\cite{48}}                                        & 66.90                      & -                            &  &  &  &  &  &  &  &  &  &  &  &  &  \\ \cline{2-4}
\multirow{-6}{*}{\textbf{SAVEE}}   & CNN {\cite{12}}                                                     & 73.6                       & -                            &  &  &  &  &  &  &  &  &  &  &  &  &  \\ \cline{1-4}
                                   & Proposed Method (Hubert)                                         & 92.82 ± 1.5                & 93.78 ± 1.6                  &  &  &  &  &  &  &  &  &  &  &  &  &  \\ \cline{2-4}
                                   & \textbf{Proposed Method (Wav2Vec2)}                              & \textbf{97.67 ± 1.2}       & \textbf{98.02 ± 1.3}         &  &  &  &  &  &  &  &  &  &  &  &  &  \\ \cline{2-4}
                                   & DCNN + CFS + MLP {\cite{48}}                                        & 73.50                      & -                            &  &  &  &  &  &  &  &  &  &  &  &  &  \\ \cline{2-4}
                                   & MFCC + Pitch + Energy + ZCR + DWT {\cite{49}}                       & -                          & 85                           &  &  &  &  &  &  &  &  &  &  &  &  &  \\ \cline{2-4}
                                   & CNN {\cite{50}}                                                     & 80                         & 79                           &  &  &  &  &  &  &  &  &  &  &  &  &  \\ \cline{2-4}
                                   & GResNets {\cite{51}} & -                          & 68.48                        &  &  &  &  &  &  &  &  &  &  &  &  &  \\ \cline{2-4}
                                   & CNN+BLSTM {\cite{52}}                                               & 69.4\%                     & 68.10                        &  &  &  &  &  &  &  &  &  &  &  &  &  \\ \cline{2-4}
\multirow{-8}{*}{\textbf{RAVDESS}} & Bagged SVM {\cite{53}}                                              & -                          & 75.69                        &  &  &  &  &  &  &  &  &  &  &  &  &  \\ \cline{1-4}
                                   & Proposed Method (Hubert)                                         & 97.83± 1.4                 & 98.21 ± 1.3                  &  &  &  &  &  &  &  &  &  &  &  &  &  \\ \cline{2-4}
                                   & \textbf{Proposed Method (Wav2Vec2)}                              & \textbf{99 ± 0.2}          & \textbf{99 ± 0.3}            &  &  &  &  &  &  &  &  &  &  &  &  &  \\ \cline{2-4}
                                   & Ensemble softmax regression {\cite{46}}                             & 82.40                      & -                            &  &  &  &  &  &  &  &  &  &  &  &  &  \\ \cline{2-4}
                                   & eGeMAPs feature set {\cite{47}}                                     & 76.90                      & -                            &  &  &  &  &  &  &  &  &  &  &  &  &  \\ \cline{2-4}
                                   & OpenSmile features + ADAN {\cite{54}}                               & 83.74                      & -                            &  &  &  &  &  &  &  &  &  &  &  &  &  \\ \cline{2-4}
                                   & RESNET MODEL + Deep BiLSTM {\cite{55}}                              & 85.57                      & -                            &  &  &  &  &  &  &  &  &  &  &  &  &  \\ \cline{2-4}
                                   & Complementary Features + KELM {\cite{56}}                           & 84.49                      & -                            &  &  &  &  &  &  &  &  &  &  &  &  &  \\ \cline{2-4}
                                   & ADRNN {\cite{57}}                                                   & 85.39                      & -                            &  &  &  &  &  &  &  &  &  &  &  &  &  \\ \cline{2-4}
                                   & DCNN + DTPM {\cite{17}}                                             & 87.31                      & -                            &  &  &  &  &  &  &  &  &  &  &  &  &  \\ \cline{2-4}
                                   & DCNN + CFS + MLP {\cite{48}}                                        & 90.50                      & -                            &  &  &  &  &  &  &  &  &  &  &  &  &  \\ \cline{2-4}
                                   & CNN + LSTM {\cite{13}}                                              & -                          & 95.89                        &  &  &  &  &  &  &  &  &  &  &  &  &  \\ \cline{2-4}
                                   & MSFs {\cite{58}}                                                    & -                          & 85.5                         &  &  &  &  &  &  &  &  &  &  &  &  &  \\ \cline{2-4}
                                   & CNN {\cite{59}}                                                     & -                          & 85.2                         &  &  &  &  &  &  &  &  &  &  &  &  &  \\ \cline{2-4}
\multirow{-14}{*}{\textbf{EMODB}}  & Fuzzy C-Means {\cite{60}}                                           & -                          & 92.2                         &  &  &  &  &  &  &  &  &  &  &  &  &  \\ \cline{1-4}
                                   & Proposed Method (Hubert)                                         & 98.36 ± 1.2                & 98.33± 1.4                   &  &  &  &  &  &  &  &  &  &  &  &  &  \\ \cline{2-4}
\multirow{-2}{*}{\textbf{AESDD}}   & Proposed Method (Wav2Vec2)                                       & 98.36 ± 1.1                & 98.33± 1.1                   &  &  &  &  &  &  &  &  &  &  &  &  &  \\ \cline{1-4}
                                   & Proposed Method (Hubert)                                         & 83.77 ± 1.7                & 71.38 ± 1.4                  &  &  &  &  &  &  &  &  &  &  &  &  &  \\ \cline{2-4}
\multirow{-2}{*}{\textbf{SHEMO}}   & Proposed Method (Wav2Vec2)                                       & \textbf{95.52± 2.1}        & \textbf{91.21± 1.3}          &  &  &  &  &  &  &  &  &  &  &  &  &  \\ \cline{1-4}
\end{tabular}
\label{tab:tbl6}
\end{table*}

\begin{table}
\caption{The confusion matrixes of SER on AESDD dataset- Wav2Vec 2.0 (left), Hubert (right).}
\centering
\resizebox{\columnwidth}{!}{%
\begin{tabular}{|c|cc|cc|cc|cc|cc|}
\hline
          & \multicolumn{2}{c|}{anger}                & \multicolumn{2}{c|}{disgust}                     & \multicolumn{2}{c|}{fear}                 & \multicolumn{2}{c|}{happiness}                   & \multicolumn{2}{c|}{sadness}                     \\ \hline
anger     & \multicolumn{1}{c|}{\textbf{100}} & 91.67 & \multicolumn{1}{c|}{0}            & 0            & \multicolumn{1}{c|}{0}     & 0            & \multicolumn{1}{c|}{0}            & 0            & \multicolumn{1}{c|} 0    & {8.33}             \\ \hline
disgust   & \multicolumn{1}{c|}{0}            & 0     & \multicolumn{1}{c|}{\textbf{100}} & \textbf{100} & \multicolumn{1}{c|}{0}     & 0            & \multicolumn{1}{c|}{0}            & 0            & \multicolumn{1}{c|}{0}            & 0            \\ \hline
fear      & \multicolumn{1}{c|}{0}            & 0     & \multicolumn{1}{c|}{0}            & 0            & \multicolumn{1}{c|}{91.67} & \textbf{100} & \multicolumn{1}{c|}{0}            & 0            & \multicolumn{1}{c|}{8.33}         & 0            \\ \hline
happiness & \multicolumn{1}{c|}{0}            & 0     & \multicolumn{1}{c|}{0}            & 0            & \multicolumn{1}{c|}{0}     & 0            & \multicolumn{1}{c|}{\textbf{100}} & \textbf{100} & \multicolumn{1}{c|}{0}            & 0            \\ \hline
sadness   & \multicolumn{1}{c|}{0}            & 0     & \multicolumn{1}{c|}{0}            & 0            & \multicolumn{1}{c|}{0}     & 0            & \multicolumn{1}{c|}{0}            & 0            & \multicolumn{1}{c|}{\textbf{100}} & \textbf{100} \\ \hline
\end{tabular}%
}
\label{tab:tbl1}
\end{table}
\begin{table}[]
\caption{The confusion matrixes of SER on EMODB dataset- Wav2Vec 2.0 (left), Hubert (right).}
\centering
\resizebox{\columnwidth}{!}{%
\begin{tabular}{|c|cc|cc|cc|cc|cc|cc|cc|}
\hline
          & \multicolumn{2}{c|}{Anger}              & \multicolumn{2}{c|}{Neutral}                     & \multicolumn{2}{c|}{Fear}               & \multicolumn{2}{c|}{Boredom}                      & \multicolumn{2}{c|}{Happiness}                   & \multicolumn{2}{c|}{Sadness}                     & \multicolumn{2}{c|}{Disgust}                     \\ \hline
Anger     & \multicolumn{1}{c|}{\textbf{100}} & \textbf{100} & \multicolumn{1}{c|}{0}            & 0            & \multicolumn{1}{c|}{0}   & 0            & \multicolumn{1}{c|}{0}            & 0             & \multicolumn{1}{c|}{0}            & 0            & \multicolumn{1}{c|}{0}            & 0            & \multicolumn{1}{c|}{0}            & 0            \\ \hline
Neutral   & \multicolumn{1}{c|}{0}            & 0   & \multicolumn{1}{c|}{\textbf{100}} & \textbf{100} & \multicolumn{1}{c|}{0}   & 0            & \multicolumn{1}{c|}{0}            & 0             & \multicolumn{1}{c|}{0}            & 0            & \multicolumn{1}{c|}{0}            & 0            & \multicolumn{1}{c|}{0}            & 0            \\ \hline
Fear      & \multicolumn{1}{c|}{0}            & 0   & \multicolumn{1}{c|}{0}            & 0            & \multicolumn{1}{c|}{\textbf{100}} & \textbf{100} & \multicolumn{1}{c|}{0}            & 0             & \multicolumn{1}{c|}{0}            & 0            & \multicolumn{1}{c|}{0}            & 0            & \multicolumn{1}{c|}{0}            & 0            \\ \hline
Boredom   & \multicolumn{1}{c|}{0}            & 0   & \multicolumn{1}{c|}{0}            & 12.5         & \multicolumn{1}{c|}{0}   & 0            & \multicolumn{1}{c|}{\textbf{100}} & {87.5} & \multicolumn{1}{c|}{0}            & 0            & \multicolumn{1}{c|}{0}            & 0            & \multicolumn{1}{c|}{0}            & 0            \\ \hline
Happiness & \multicolumn{1}{c|}{0}            & 0   & \multicolumn{1}{c|}{0}            & 0            & \multicolumn{1}{c|}{0}   & 0            & \multicolumn{1}{c|}{0}            & 0             & \multicolumn{1}{c|}{\textbf{100}} & \textbf{100} & \multicolumn{1}{c|}{0}            & 0            & \multicolumn{1}{c|}{0}            & 0            \\ \hline
Sadness   & \multicolumn{1}{c|}{0}            & 0   & \multicolumn{1}{c|}{0}            & 0            & \multicolumn{1}{c|}{0}   & 0            & \multicolumn{1}{c|}{0}            & 0             & \multicolumn{1}{c|}{0}            & 0            & \multicolumn{1}{c|}{\textbf{100}} & \textbf{100} & \multicolumn{1}{c|}{0}            & 0            \\ \hline
Disgust   & \multicolumn{1}{c|}{0}            & 0   & \multicolumn{1}{c|}{0}            & 0            & \multicolumn{1}{c|}{0}   & 0            & \multicolumn{1}{c|}{0}            & 0             & \multicolumn{1}{c|}{0}            & 0            & \multicolumn{1}{c|}{0}            & 0            & \multicolumn{1}{c|}{\textbf{100}} & \textbf{100} \\ \hline
\end{tabular}
}
\label{tab:tbl2}
\end{table}

\begin{table}[]
\caption{The confusion matrixes of SER on RAVDESS dataset- Wav2Vec 2.0 (left), Hubert (right).}
\centering
\resizebox{\columnwidth}{!}{%
\begin{tabular}{|c|cc|cc|cc|cc|cc|cc|cc|cc|}
\hline
        & \multicolumn{2}{c|}{Neutral}              & \multicolumn{2}{c|}{Calm}                 & \multicolumn{2}{c|}{Happy}                  & \multicolumn{2}{c|}{Sad}                    & \multicolumn{2}{c|}{Angry}                  & \multicolumn{2}{c|}{Fearful}              & \multicolumn{2}{c|}{Disgust}              & \multicolumn{2}{c|}{Wonder}               \\ \hline
Neutral & \multicolumn{1}{c|}{94.74} & \textbf{100} & \multicolumn{1}{c|}{5.26}         & 0     & \multicolumn{1}{c|}{0}              & 0     & \multicolumn{1}{c|}{0}              & 0     & \multicolumn{1}{c|}{0}              & 0     & \multicolumn{1}{c|}{0}            & 0     & \multicolumn{1}{c|}{0}            & 0     & \multicolumn{1}{c|}{0}            & 0     \\ \hline
Calm    & \multicolumn{1}{c|}{0}     & 5.13         & \multicolumn{1}{c|}{\textbf{100}} & 94.87 & \multicolumn{1}{c|}{0}              & 0     & \multicolumn{1}{c|}{0}              & 0     & \multicolumn{1}{c|}{0}              & 0     & \multicolumn{1}{c|}{0}            & 0     & \multicolumn{1}{c|}{0}            & 0     & \multicolumn{1}{c|}{0}            & 0     \\ \hline
Happy   & \multicolumn{1}{c|}{0}     & 0            & \multicolumn{1}{c|}{0}            & 0     & \multicolumn{1}{c|}{\textbf{97.37}} & 86.84 & \multicolumn{1}{c|}{0}              & 2.63  & \multicolumn{1}{c|}{0}              & 0     & \multicolumn{1}{c|}{0}            & 2.63  & \multicolumn{1}{c|}{0}            & 0     & \multicolumn{1}{c|}{2.63}         & 7.89  \\ \hline
Sad     & \multicolumn{1}{c|}{2.63}  & 2.63         & \multicolumn{1}{c|}{0}            & 0     & \multicolumn{1}{c|}{2.63}           & 2.63  & \multicolumn{1}{c|}{\textbf{94.74}} & 86.84 & \multicolumn{1}{c|}{0}              & 0     & \multicolumn{1}{c|}{0}            & 7.89  & \multicolumn{1}{c|}{0}            & 0     & \multicolumn{1}{c|}{0}            & 0     \\ \hline
Angry   & \multicolumn{1}{c|}{0}     & 2.63         & \multicolumn{1}{c|}{0}            & 0     & \multicolumn{1}{c|}{0}              & 0     & \multicolumn{1}{c|}{0}              & 0     & \multicolumn{1}{c|}{\textbf{97.37}} & 89.48 & \multicolumn{1}{c|}{0}            & 0     & \multicolumn{1}{c|}{0}            & 0     & \multicolumn{1}{c|}{2.63}         & 7.89  \\ \hline
Fearful & \multicolumn{1}{c|}{0}     & 0            & \multicolumn{1}{c|}{0}            & 0     & \multicolumn{1}{c|}{0}              & 2.56  & \multicolumn{1}{c|}{0}              & 0     & \multicolumn{1}{c|}{0}              & 0     & \multicolumn{1}{c|}{\textbf{100}} & 97.44 & \multicolumn{1}{c|}{0}            & 0     & \multicolumn{1}{c|}{0}            & 0     \\ \hline
Disgust & \multicolumn{1}{c|}{0}     & 0            & \multicolumn{1}{c|}{0}            & 0     & \multicolumn{1}{c|}{0}              & 0     & \multicolumn{1}{c|}{0}              & 0     & \multicolumn{1}{c|}{0}              & 2.63  & \multicolumn{1}{c|}{0}            & 0     & \multicolumn{1}{c|}{\textbf{100}} & 97.37 & \multicolumn{1}{c|}{0}            & 0     \\ \hline
Wonder  & \multicolumn{1}{c|}{0}     & 0            & \multicolumn{1}{c|}{0}            & 0     & \multicolumn{1}{c|}{0}              & 2.56  & \multicolumn{1}{c|}{0}              & 0     & \multicolumn{1}{c|}{0}              & 0     & \multicolumn{1}{c|}{0}            & 0     & \multicolumn{1}{c|}{0}            & 0     & \multicolumn{1}{c|}{\textbf{100}} & 97.44 \\ \hline
\end{tabular}
}
\label{tab:tbl3}
\end{table}

\begin{table}[]
\caption{The confusion matrixes of SER on SAVEE dataset- Wav2Vec 2.0 (left), Hubert (right).}
\centering
\resizebox{\columnwidth}{!}{%
\begin{tabular}{cccccccccccccccll}
\cline{1-15}
\multicolumn{1}{|c|}{}          & \multicolumn{2}{c|}{Anger}                                            & \multicolumn{2}{c|}{Disgust}                                    & \multicolumn{2}{c|}{Fear}                                   & \multicolumn{2}{c|}{Happiness}                                   & \multicolumn{2}{c|}{Sadness}                                   & \multicolumn{2}{c|}{Wonder}                                  & \multicolumn{2}{c|}{Neutral}                                          &  &  \\ \cline{1-15}
\multicolumn{1}{|c|}{Anger}     & \multicolumn{1}{c|}{\textbf{100}} & \multicolumn{1}{c|}{\textbf{100}} & \multicolumn{1}{c|}{0}              & \multicolumn{1}{c|}{0}    & \multicolumn{1}{c|}{0}            & \multicolumn{1}{c|}{0}  & \multicolumn{1}{c|}{0}     & \multicolumn{1}{c|}{0}              & \multicolumn{1}{c|}{0}     & \multicolumn{1}{c|}{0}            & \multicolumn{1}{c|}{0} & \multicolumn{1}{c|}{0}              & \multicolumn{1}{c|}{0}            & \multicolumn{1}{c|}{0}            &  &  \\ \cline{1-15}
\multicolumn{1}{|c|}{Disgust}   & \multicolumn{1}{c|}{8.33}         & \multicolumn{1}{c|}{8.33}         & \multicolumn{1}{c|}{\textbf{91.67}} & \multicolumn{1}{c|}{75.0} & \multicolumn{1}{c|}{0}            & \multicolumn{1}{c|}{0}  & \multicolumn{1}{c|}{0}     & \multicolumn{1}{c|}{0}              & \multicolumn{1}{c|}{0}     & \multicolumn{1}{c|}{8.33}         & \multicolumn{1}{c|}{0} & \multicolumn{1}{c|}{8.34}           & \multicolumn{1}{c|}{0}            & \multicolumn{1}{c|}{0}            &  &  \\ \cline{1-15}
\multicolumn{1}{|c|}{Fear}      & \multicolumn{1}{c|}{0}            & \multicolumn{1}{c|}{0}            & \multicolumn{1}{c|}{0}              & \multicolumn{1}{c|}{0}    & \multicolumn{1}{c|}{\textbf{100}} & \multicolumn{1}{c|}{75} & \multicolumn{1}{c|}{0}     & \multicolumn{1}{c|}{16.67}          & \multicolumn{1}{c|}{8}     & \multicolumn{1}{c|}{0}            & \multicolumn{1}{c|}{0} & \multicolumn{1}{c|}{8.33}           & \multicolumn{1}{c|}{0}            & \multicolumn{1}{c|}{0}            &  &  \\ \cline{1-15}
\multicolumn{1}{|c|}{Happiness} & \multicolumn{1}{c|}{8.33}         & \multicolumn{1}{c|}{8.33}         & \multicolumn{1}{c|}{8.33}           & \multicolumn{1}{c|}{0}    & \multicolumn{1}{c|}{0}            & \multicolumn{1}{c|}{0}  & \multicolumn{1}{c|}{83.34} & \multicolumn{1}{c|}{\textbf{91.67}} & \multicolumn{1}{c|}{0}     & \multicolumn{1}{c|}{0}            & \multicolumn{1}{c|}{0} & \multicolumn{1}{c|}{0}              & \multicolumn{1}{c|}{0}            & \multicolumn{1}{c|}{0}            &  &  \\ \cline{1-15}
\multicolumn{1}{|c|}{Sadness}   & \multicolumn{1}{c|}{0}            & \multicolumn{1}{c|}{0}            & \multicolumn{1}{c|}{0}              & \multicolumn{1}{c|}{8}    & \multicolumn{1}{c|}{8.33}         & \multicolumn{1}{c|}{0}  & \multicolumn{1}{c|}{0}     & \multicolumn{1}{c|}{0}              & \multicolumn{1}{c|}{91.67} & \multicolumn{1}{c|}{\textbf{100}} & \multicolumn{1}{c|}{0} & \multicolumn{1}{c|}{0}              & \multicolumn{1}{c|}{0}            & \multicolumn{1}{c|}{0}            &  &  \\ \cline{1-15}
\multicolumn{1}{|c|}{Wonder}    & \multicolumn{1}{c|}{0}            & \multicolumn{1}{c|}{8.33}         & \multicolumn{1}{c|}{0}              & \multicolumn{1}{c|}{0}    & \multicolumn{1}{c|}{83.33}        & \multicolumn{1}{c|}{0}  & \multicolumn{1}{c|}{16.67} & \multicolumn{1}{c|}{0}              & \multicolumn{1}{c|}{0}     & \multicolumn{1}{c|}{0}            & \multicolumn{1}{c|}{0} & \multicolumn{1}{c|}{\textbf{91.67}} & \multicolumn{1}{c|}{0}            & \multicolumn{1}{c|}{0}            &  &  \\ \cline{1-15}
\multicolumn{1}{|c|}{Neutral}   & \multicolumn{1}{c|}{0}            & \multicolumn{1}{c|}{0}            & \multicolumn{1}{c|}{0}              & \multicolumn{1}{c|}{0}    & \multicolumn{1}{c|}{0}            & \multicolumn{1}{c|}{0}  & \multicolumn{1}{c|}{0}     & \multicolumn{1}{c|}{0}              & \multicolumn{1}{c|}{0}     & \multicolumn{1}{c|}{0}            & \multicolumn{1}{c|}{0} & \multicolumn{1}{c|}{0}              & \multicolumn{1}{c|}{\textbf{100}} & \multicolumn{1}{c|}{\textbf{100}} &  &  \\ \cline{1-15}
\multicolumn{1}{l}{}            & \multicolumn{1}{l}{}              & \multicolumn{1}{l}{}              & \multicolumn{1}{l}{}                & \multicolumn{1}{l}{}      & \multicolumn{1}{l}{}              & \multicolumn{1}{l}{}    & \multicolumn{1}{l}{}       & \multicolumn{1}{l}{}                & \multicolumn{1}{l}{}       & \multicolumn{1}{l}{}              & \multicolumn{1}{l}{}   & \multicolumn{1}{l}{}                & \multicolumn{1}{l}{}              & \multicolumn{1}{l}{}              &  & 
\end{tabular}
}
\label{tab:tbl4}
\end{table}

\begin{table}
\caption{The confusion matrixes of SER on SHEMO dataset- Wav2Vec 2.0 (left), Hubert (right).}
\centering
\resizebox{1.2\columnwidth}{!}{%
\begin{tabular}{cccccccccccllllll}
\cline{1-11}
\multicolumn{1}{|c|}{}          & \multicolumn{2}{c|}{Anger}                                       & \multicolumn{2}{c|}{Happiness}                                  & \multicolumn{2}{c|}{Sadness}                                     & \multicolumn{2}{c|}{Surprise}                                    & \multicolumn{2}{c|}{Neutral}                                     &  &  &  &  &  &  \\ \cline{1-11}
\multicolumn{1}{|c|}{Anger}     & \multicolumn{1}{c|}{\textbf{99.06}} & \multicolumn{1}{c|}{92.45} & \multicolumn{1}{c|}{0.94}           & \multicolumn{1}{c|}{1.89} & \multicolumn{1}{c|}{0}              & \multicolumn{1}{c|}{0}     & \multicolumn{1}{c|}{0}              & \multicolumn{1}{c|}{0.94}  & \multicolumn{1}{c|}{0}              & \multicolumn{1}{c|}{4.72}  &  &  &  &  &  &  \\ \cline{1-11}
\multicolumn{1}{|c|}{Happiness} & \multicolumn{1}{c|}{0}              & \multicolumn{1}{c|}{27.5}  & \multicolumn{1}{c|}{\textbf{87.50}} & \multicolumn{1}{c|}{42.5} & \multicolumn{1}{c|}{5.0}            & \multicolumn{1}{c|}{12.5}  & \multicolumn{1}{c|}{0}              & \multicolumn{1}{c|}{2.5}   & \multicolumn{1}{c|}{7.5}            & \multicolumn{1}{c|}{15.0}  &  &  &  &  &  &  \\ \cline{1-11}
\multicolumn{1}{|c|}{Sadness}   & \multicolumn{1}{c|}{0}              & \multicolumn{1}{c|}{2.22}  & \multicolumn{1}{c|}{0}              & \multicolumn{1}{c|}{5.56} & \multicolumn{1}{c|}{\textbf{92.22}} & \multicolumn{1}{c|}{74.44} & \multicolumn{1}{c|}{0}              & \multicolumn{1}{c|}{2.22}  & \multicolumn{1}{c|}{7.78}           & \multicolumn{1}{c|}{15.56} &  &  &  &  &  &  \\ \cline{1-11}
\multicolumn{1}{|c|}{Surprise}  & \multicolumn{1}{c|}{0}              & \multicolumn{1}{c|}{6.67}  & \multicolumn{1}{c|}{0}              & \multicolumn{1}{c|}{6.67} & \multicolumn{1}{c|}{0}              & \multicolumn{1}{c|}{4.44}  & \multicolumn{1}{c|}{\textbf{77.78}} & \multicolumn{1}{c|}{53.33} & \multicolumn{1}{c|}{22.22}          & \multicolumn{1}{c|}{28.89} &  &  &  &  &  &  \\ \cline{1-11}
\multicolumn{1}{|c|}{Neutral}   & \multicolumn{1}{c|}{0}              & \multicolumn{1}{c|}{3.88}  & \multicolumn{1}{c|}{0}              & \multicolumn{1}{c|}{0}    & \multicolumn{1}{c|}{0.49}           & \multicolumn{1}{c|}{0.97}  & \multicolumn{1}{c|}{0}              & \multicolumn{1}{c|}{0.97}  & \multicolumn{1}{c|}{\textbf{99.51}} & \multicolumn{1}{c|}{94.18} &  &  &  &  &  &  \\ \cline{1-11}
\multicolumn{1}{l}{}            & \multicolumn{1}{l}{}                & \multicolumn{1}{l}{}       & \multicolumn{1}{l}{}                & \multicolumn{1}{l}{}      & \multicolumn{1}{l}{}                & \multicolumn{1}{l}{}       & \multicolumn{1}{l}{}                & \multicolumn{1}{l}{}       & \multicolumn{1}{l}{}                & \multicolumn{1}{l}{}       &  &  &  &  &  &  \\
\multicolumn{1}{l}{}            & \multicolumn{1}{l}{}                & \multicolumn{1}{l}{}       & \multicolumn{1}{l}{}                & \multicolumn{1}{l}{}      & \multicolumn{1}{l}{}                & \multicolumn{1}{l}{}       & \multicolumn{1}{l}{}                & \multicolumn{1}{l}{}       & \multicolumn{1}{l}{}                & \multicolumn{1}{l}{}       &  &  &  &  &  &  \\
\multicolumn{1}{l}{}            & \multicolumn{1}{l}{}                & \multicolumn{1}{l}{}       & \multicolumn{1}{l}{}                & \multicolumn{1}{l}{}      & \multicolumn{1}{l}{}                & \multicolumn{1}{l}{}       & \multicolumn{1}{l}{}                & \multicolumn{1}{l}{}       & \multicolumn{1}{l}{}                & \multicolumn{1}{l}{}       &  &  &  &  &  & 
\end{tabular}
}
\label{tab:tbl5}
\end{table}

Several experiments were carried out on the explained dataset by the proposed model. We split the datasets into 80\% and 20\% for train and test respectively, we kept the ratio of the classes in the test set. The confusion matrix is one of the best methods for showing the evaluation of the methods for the SER problem since we can analyze the missed classifications. For this study, we applied large models of HUBERT and Wave2Vec 2.0.

Our work aims to recognize speech emotion with high generalization performance and high accuracy. We used transformer-based models for feature extraction. Tables  \ref{tab:tbl1}--\ref{tab:tbl5} show that our proposed SER model performed acceptably across speakers with different languages and speaking styles. In table \ref{tab:tbl6}, we present the results of testing our methods on five folds, comparing both weighted and unweighted accuracy with well-known models. As shown, the proposed models outperformed various methods in emotion recognition.

This section provides a detailed analysis of certain misclassifications observed in the tables. In table \ref{tab:tbl2}, the Hubert-based model misclassified boredom as neutral due to the similar tone of voice in the test data. Similarly, in table \ref{tab:tbl3}, anger was frequently misclassified as wonder, and calm as neutral, likely due to the tonal similarities between these classes. The Hubert-based model misclassified happy voices as wonder. After listening to the samples, we observed a similarity between the tones of happy and wonder emotions. In the samples that were sad but recognized as fearful, the tone of the speakers closely resembled that of the fearful emotion.

In table \ref{tab:tbl4}, the Wav2Vec2-based model struggled with the wonder class; upon examining the score of the incorrect class (fear), we found it to be close to 0.60, indicating that the model might benefit from additional training data. Lastly, in table \ref{tab:tbl5}, both models misclassified wonder from neutral. After analyzing the common misclassified samples, we concluded that the tone in the wonder class closely resembled neutral and that these samples tended to have shorter durations. Furthermore, the Hubert-based model misclassified some happy samples as neutral. After listening to these samples, we found a high degree of similarity between the tones of happy and neutral speech.

\section{Conclusion}
This paper proposes transformer-based feature extractors as a way of solving the SER problem. Using transformer-based models, we can learn many useful temporal and spatial features from raw audio files that are very useful for emotion recognition. Our claim is supported by experimental results from benchmark datasets. The two models performed well on standard and famous datasets, and their performance was acceptable. This shows the significance and effectiveness of the proposed system for SER using self-supervised methods. Moreover, our model performed well on noisy data such as that from call centers, which is relevant to real-world applications.
Although the self-supervised models presented in this paper have demonstrated improved performance in speaker emotion recognition, there are still opportunities for further enhancing the model. We can use CNN models in the following of the output of the feature extraction part. In this case, we calculated the mean of the features. By using CNN models we can capture a more meaningful relationship between the features rather than a mean measurement. Also, performance could be further improved by augmenting the training dataset with noise, SpecAugment, and room impulse responses (RIRs).

\bibliographystyle{plain}
\bibliography{main}

\begin{thebibliography}{10}

\bibitem{7}
J.~P. Arias, B.~Carlos, and B.~Nestor.
\newblock Shape-based modeling of the fundamental frequency contour for emotion detection in speech.
\newblock {\em Computer Speech and Language}, 28(1):278--294, 2014.

\bibitem{14}
A.~M. Badshah and et~al.
\newblock Speech emotion recognition from spectrograms with deep convolutional neural network.
\newblock In {\em 2017 International Conference on Platform Technology and Service (PlatCon)}. IEEE, 2017.

\bibitem{11}
A.~M. Badshah and et~al.
\newblock Deep features-based speech emotion recognition for smart affective services.
\newblock {\em Multimedia Tools and Applications}, 78:5571--5589, 2019.

\bibitem{26}
A.~Baevski and et~al.
\newblock wav2vec 2.0: A framework for self-supervised learning of speech representations.
\newblock In {\em Advances in Neural Information Processing Systems}, volume~33, pages 12449--12460, 2020.

\bibitem{38}
A.~Baevski, S.~Schneider, and M.~Au.
\newblock vq-wav2vec: Self-supervised learning of discrete speech representations.
\newblock {\em arXiv preprint arXiv:1910.05453}, 2019.

\bibitem{53}
A.~Bhavan, P.~Chauhan, and R.~R. Sha.
\newblock Bagged support vector machines for emotion recognition from speech.
\newblock {\em Knowledge-Based Systems}, 184:104886, 2019.

\bibitem{43}
F.~Burkhardt and et~al.
\newblock A database of german emotional speech.
\newblock In {\em Interspeech}, volume~5, 2005.

\bibitem{28}
Y.~Chavhan, M.~L. Dhore, and P.~Yes.
\newblock Speech emotion recognition using support vector machine.
\newblock {\em International Journal of Computer Applications}, 1:6--9, 2010.

\bibitem{21}
M.~Chen and et~al.
\newblock 3-d convolutional recurrent neural networks with attention model for speech emotion recognition.
\newblock {\em IEEE Signal Processing Letters}, 25:1440--1444, 2018.

\bibitem{18}
J.~K. Chorowsk and et~al.
\newblock Attention-based models for speech recognition.
\newblock In {\em Advances in Neural Information Processing Systems}, volume~28, 2015.

\bibitem{1}
R.~Cowie, E.~Douglas-Cowie, N.~Tsapatsoulis, G.~Votsis, S.~Kollias, W.~Fellenz, and J.~Taylor.
\newblock Emotion recognition in human-computer interaction.
\newblock {\em IEEE Signal Processing Magazine}, 18:32--80, 2001.

\bibitem{60}
S.~Demircan and H.~Kahramanl.
\newblock Application of fuzzy c-means clustering algorithm to spectral features for emotion classification from speech.
\newblock {\em Neural Computing and Applications}, 29:59--66, 2018.

\bibitem{48}
M.~Farooq and et~al.
\newblock Impact of feature selection algorithm on speech emotion recognition using deep convolutional neural network.
\newblock {\em Sensors}, 21:20, 2020.

\bibitem{56}
L.~Guo and et~al.
\newblock Exploration of complementary features for speech emotion recognition based on kernel extreme learning machine.
\newblock {\em IEEE Access}, 7:75798--75809, 2019.

\bibitem{47}
F.~Haider and et~al.
\newblock Emotion recognition in low-resource settings: An evaluation of automatic feature selection methods.
\newblock {\em Computer Speech and Language}, 65, 2021.

\bibitem{37}
D.~Hendrycks and K.~Gimpel.
\newblock Gaussian error linear units (gelus).
\newblock {\em arXiv preprint arXiv:1606.08415}, 2016.

\bibitem{39}
J.~Herve, M.~Douze, and C.~Schmid.
\newblock Product quantization for nearest neighbor search.
\newblock {\em IEEE Transactions on Pattern Analysis and Machine Intelligence}, 33(1):117--128, 2010.

\bibitem{32}
G.~A.~Ten Holt, M.~J. Reinders, and E.~A. Hendriks.
\newblock Multi-dimensional dynamic time warping for gesture recognition.
\newblock In {\em Thirteenth Annual Conference of the Advanced School for Computing and Imaging}, 2007.

\bibitem{25}
W.-N. Hsu and et~al.
\newblock Hubert: Self-supervised speech representation learning by masked prediction of hidden units.
\newblock {\em IEEE/ACM Transactions on Audio, Speech, and Language Processing}, 29:3451--3460, 2021.

\bibitem{27}
H.~Hu, X.~Ming-Xing, and W.~Wu.
\newblock Gmm supervector based svm with spectral features for speech emotion recognition.
\newblock In {\em 2007 IEEE International Conference on Acoustics, Speech and Signal Processing - ICASSP'07}, 2007.

\bibitem{23}
C.-W. Huang and S.~S. Narayanan.
\newblock Attention assisted discovery of sub-utterance structure in speech emotion recognition.
\newblock In {\em Interspeech}, 2016.

\bibitem{24}
C.-W. Huang and S.~S. Narayanan.
\newblock Deep convolutional recurrent neural network with attention mechanism for robust speech emotion recognition.
\newblock In {\em 2017 IEEE International Conference on Multimedia and Expo (ICME)}. IEEE, 2017.

\bibitem{44}
P.~Jackson and S.~Haq.
\newblock Surrey audio-visual expressed emotion (savee) database.
\newblock In {\em University of Surrey: Guildford}, 2014.

\bibitem{52}
M.~A. Jalal and et~al.
\newblock Learning temporal clusters using capsule routing for speech emotion recognition.
\newblock In {\em Proceedings of Interspeech}, 2019.

\bibitem{36}
M.~Joshi and et~al.
\newblock Spanbert: Improving pre-training by representing and predicting spans.
\newblock {\em Transactions of the Association for Computational Linguistics}, 8:64--77, 2020.

\bibitem{35}
J.~Kahn and et~al.
\newblock Libri-light: A benchmark for asr with limited or no supervision.
\newblock In {\em ICASSP 2020 - IEEE International Conference on Acoustics, Speech and Signal Processing (ICASSP)}. IEEE, 2020.

\bibitem{30}
Y.~Kim and E.~Mower Provost.
\newblock Emotion classification via utterance-level dynamics: A pattern-based approach to characterizing affective expressions.
\newblock In {\em 2013 IEEE International Conference on Acoustics, Speech and Signal Processing (ICASSP)}. IEEE, 2013.

\bibitem{49}
A.~Koduru, H.~B. Valiveti, and A.~Kumar Budati.
\newblock Feature extraction algorithms to improve the speech emotion recognition rate.
\newblock {\em International Journal of Speech Technology}, 23(1):45--55, 2020.

\bibitem{31}
C.-C. Lee and et~al.
\newblock Emotion recognition using a hierarchical binary decision tree approach.
\newblock {\em Speech Communication}, 53:1162--1171, 2011.

\bibitem{16}
W.~Lim, J.~Daeyoung, and L.~Taejin.
\newblock Speech emotion recognition using convolutional and recurrent neural networks.
\newblock In {\em Asia-Pacific Signal and Information Processing Association Annual Summit and Conference (APSIPA)}. IEEE, 2016.

\bibitem{40}
S.~R. Livingstone and F.~A. Russo.
\newblock The ryerson audio-visual database of emotional speech and song (ravdess): A dynamic, multimodal set of facial and vocal expressions in north american english.
\newblock {\em PloS One}, 13(5), 2018.

\bibitem{3}
B.~Maier and W.~A. Shibles.
\newblock Emotion in medicine.
\newblock In {\em The Philosophy and Practice of Medicine and Bioethics: A Naturalistic-Humanistic Approach}, pages 137--159. Springer, Dordrecht, 2011.

\bibitem{12}
Q.~Mao and et~al.
\newblock Learning salient features for speech emotion recognition using convolutional neural networks.
\newblock {\em IEEE Transactions on Multimedia}, 16:2203--2213, 2014.

\bibitem{57}
H.~Meng and et~al.
\newblock Speech emotion recognition from 3d log-mel spectrograms with deep learning network.
\newblock {\em IEEE Access}, 7:125868--125881, 2019.

\bibitem{22}
S.~Mirsamadi, B.~Emad, and Z.~Cha.
\newblock Automatic speech emotion recognition using recurrent neural networks with local attention.
\newblock In {\em 2017 IEEE International Conference on Acoustics, Speech and Signal Processing (ICASSP)}. IEEE, 2017.

\bibitem{19}
V.~Mnih, N.~Heess, and A.~Graves.
\newblock Recurrent models of visual attention.
\newblock In {\em Advances in Neural Information Processing Systems}, volume~27, 2014.

\bibitem{50}
Mustaqeem and S.~Kwon.
\newblock A cnn-assisted enhanced audio signal processing for speech emotion recognition.
\newblock {\em Sensors}, 20(1):183, 2019.

\bibitem{45}
O.~M. Nezami, P.~Jamshid Lou, and M.~Karami.
\newblock Shemo: A large-scale validated database for persian speech emotion detection.
\newblock {\em Language Resources and Evaluation}, 53:1--16, 2019.

\bibitem{34}
V.~Panayotov and et~al.
\newblock Librispeech: An asr corpus based on public domain audio books.
\newblock In {\em IEEE International Conference on Acoustics, Speech and Signal Processing (ICASSP)}. IEEE, 2015.

\bibitem{9}
T.-L. Pao and et~al.
\newblock Mandarin emotional speech recognition based on svm and nn.
\newblock In {\em 18th International Conference on Pattern Recognition (ICPR)}, volume~1. IEEE, 2016.

\bibitem{2}
S.~Ramakrishnan and M.~E.~E. Ibrahiem.
\newblock Speech emotion recognition approaches in human computer interaction.
\newblock {\em Telecommunication Systems}, 52:1467--1478, 2013.

\bibitem{55}
M.~Sajjad and S.~Kwon.
\newblock Clustering-based speech emotion recognition by incorporating learned features.
\newblock {\em IEEE Access}, 8:79861--79875, 2020.

\bibitem{6}
B.~Schuller, A.~Dejan, W.~Frank, and R.~Gerhard.
\newblock Emotion recognition in the noise applying large acoustic feature sets.
\newblock In {\em Proceedings of the 2006 IEEE International Conference}, 2006.

\bibitem{8}
B.~Schuller, R.~Gerhard, and L.~Manfred.
\newblock Hidden markov model-based speech emotion recognition.
\newblock In {\em 2003 IEEE International Conference on Acoustics, Speech, and Signal Processing, ICASSP}, 2003.

\bibitem{29}
S.~S. Shambhavi and V.~N. Nitnaware.
\newblock Emotion speech recognition using mfcc and svm.
\newblock {\em International Journal of Engineering Research and Technology}, 4:1067--1070, 2015.

\bibitem{10}
P.~Shen, Z.~Changjun, and X.~Chen.
\newblock Automatic speech emotion recognition using support vector machine.
\newblock In {\em Proceedings of the 2011 International Conference on Electronic and Mechanical Engineering and Information Technology}, 2011.

\bibitem{33}
C.~Subakan and et~al.
\newblock Attention is all you need in speech separation.
\newblock In {\em ICASSP 2021 - IEEE International Conference on Acoustics, Speech and Signal Processing (ICASSP)}, 2021.

\bibitem{46}
Y.~Sun and G.~Wen.
\newblock Ensemble softmax regression model for speech emotion recognition.
\newblock {\em Multimedia Tools and Applications}, 76:8305--8328, 2017.

\bibitem{41}
N.~Vryzas and et~al.
\newblock Speech emotion recognition for performance interaction.
\newblock {\em Journal of the Audio Engineering Society}, 66(6):457--467, 2018.

\bibitem{42}
N.~Vryzas and et~al.
\newblock Subjective evaluation of a speech emotion recognition interaction framework.
\newblock In {\em Proceedings of the Audio Mostly 2018 on Sound in Immersion and Emotion}, pages 1--7, 2018.

\bibitem{5}
K.~Wang, A.~Ning, N.~L. Bing, and Z.~Yanyong.
\newblock Speech emotion recognition using fourier parameters ieee transactions on affective computing.
\newblock {\em IEEE Transactions on Affective Computing}, 6:69--75, 2015.

\bibitem{58}
S.~Wu, H.~F. Tiago, and W.-Y. Chan.
\newblock Automatic speech emotion recognition using modulation spectral features.
\newblock {\em Speech Communication}, 53(5):768--785, 2011.

\bibitem{20}
Z.~Yang and et~al.
\newblock Hierarchical attention networks for document classification.
\newblock In {\em Proceedings of the 2016 Conference of the North American Chapter of the Association for Computational Linguistics: Human Language Technologies}, 2016.

\bibitem{54}
L.~Yi and M.~Man-Wai.
\newblock Adversarial data augmentation network for speech emotion recognition.
\newblock In {\em 2019 Asia-Pacific Signal and Information Processing Association Annual Summit and Conference (APSIPA ASC)}. IEEE, 2019.

\bibitem{51}
Y.~Zeng and et~al.
\newblock Spectrogram based multi-task audio classification.
\newblock {\em Multimedia Tools and Applications}, 78:3705--3722, 2019.

\bibitem{17}
S.~Zhang and et~al.
\newblock Speech emotion recognition using deep convolutional neural network and discriminant temporal pyramid matching.
\newblock {\em IEEE Transactions on Multimedia}, 20:1576--1590, 2017.

\bibitem{13}
J.~Zhao, M.~Xia, and C.~Lijiang.
\newblock Speech emotion recognition using deep 1d and 2d cnn lstm networks.
\newblock {\em Biomedical Signal Processing and Control}, 47:312--323, 2019.

\bibitem{15}
W.~Q. Zheng, J.~S. Yu, and Y.~X. Zou.
\newblock An experimental study of speech emotion recognition based on deep convolutional neural networks.
\newblock In {\em 2015 International Conference on Affective Computing and Intelligent Interaction (ACII)}. IEEE, 2015.

\bibitem{59}
H.~Zhengwei, D.~Ming, M.~Qirong, and Z.~Yongzhao.
\newblock Speech emotion recognition using cnn.
\newblock In {\em Association for Computing Machinery}, volume~4, pages 801--804, 2014.

\bibitem{4}
D.~C. Zuroff and A.~C. Sally.
\newblock Emotion recognition in schizophrenic and depressed inpatients.
\newblock {\em Journal of Clinical Psychology}, 42(3):411--417, 1986.

\end{thebibliography}
\end{document}